\documentclass[english]{article}
\usepackage[T1]{fontenc}
\usepackage[latin1]{inputenc}
\usepackage{geometry}
\usepackage{amsmath}
\usepackage{setspace}
\usepackage{amssymb}

\makeatletter
\newcommand{\lyxaddress}[1]{
\par {\raggedright #1
\vspace{1.4em}
\noindent\par}
}

\usepackage{babel}
\makeatother
\begin{document}

\title{INTERPRETATIONS OF OCTONION WAVE EQUATIONS}

\author{P. S. Bisht$^{\text{(1)}}$ , Bimal Pandey$^{\text{(1)}}$ and O.
P. S. Negi$^{\text{(2)}}$ %
\thanks{Permanent Address- Department of Physics, Kumaun University, S. S.
J. Campus, Almora -263601 (U.A.) INDIA%
}}

\maketitle

\lyxaddress{\begin{center}
$^{\text{(1)}}$Department of Physics,\\
 Kumaun University, \\
S. S. J. Campus,\\
 Almora - 263601 (Uttarakhand) India
\par\end{center}}

\lyxaddress{\begin{center}
$^{\text{(2)}}$Institute of Theoretical Physics,\\
 Chinese Academy of Sciences\\
 KITP Building Room No.- 6304\\
Hai Dian Qu Zhong Guan Chun Dong Lu\\
 55 Hao , Beijing - 100080, P.R.China 
\par\end{center}}

\lyxaddress{\begin{center}
Email - ps\_bisht123@ rediffmail.com,\\
ops\_negi @ yahoo.co.in
\par\end{center}}

\begin{abstract}
The interpretations of octonion wave equations in eight dimensional
space-time have been discussed. We have made an attempt to discuss
the octonion field equation as the equation of motion for particles
carrying simultaneously electric and magnetic charges (i.e. dyons)
in external and internal spaces respectively. It has been concluded
that the component of octonion potential wave equations behaves neither
the generalized electromagnetic fields of monopoles nor the dyons.
Rather, they have the mixed behaviour of electromagnetic fields associated
with the electric and magnetic charges in external and internal spaces.
We have also made an attempt to investigate the split octonion wave
equation and its interpretation in classical electrodynamics and accordingly
the consistent and compact forms of eight-dimensional potential and
current equation of dyons are obtained in terms of Zorn's vector matrix
realization of split octonions.Visualizing the external four space
as the localization space for tachyons, it is shown that the split
octonion wave equation reduces to the Maxwell's equation (field equation)
for bradyons in $R^{4}$- space as well as that for tachyons in $T^{4}$-
space in the absence of other.
\end{abstract}

\section{INTRODUCTION}

Octonions were first introduced in Physics by Jordon, Von Neuman and
Wigner\cite{key-1}, who investigated a new finite Hilbert space,
on replacing the complex numbers by Octonions\cite{key-2}. According
to celebrated Hurwitz theorem \cite{key-3} there exits four-division
algebra consisting of $\mathbb{R}$ (real numbers), $\mathbb{C}$
(complex numbers), $\mathbb{H}$ (quaternions) and $\mathcal{O}$
(octonions). All four algebras are alternative with antisymmetric
associators. In 1961 Pais \cite{key-4} pointed out a striking similarity
between the algebra of interactions and the split octonion algebra.
Recently, some attention has been directed to octonions in theoretical
physics with the hope of extending the 3+1 space-time framework of
the theory to eight dimensions to accommodate the ever increasing
quantum numbers and internal symmetries assigned to elementary particles
and gauge fields. A lot of literature is available \cite{key-5,key-6,key-7,key-8,key-9,key-10,key-11}
on the applications of octonions to interpret wave equation, Dirac
equation, and the extension of octonion non-associativity to physical
theories. It is assumed there that octonion wave equation contains
four-dimensional internal space of mass parameters while the external
four dimensional space has been interpreted as the four generators
of fermions within the quark-lepton symmetry. The ingenious work,
in this direction is done by Günaydin and Gürsey \cite{key-12,key-13},
to formulate quark models and colour gauge theory in terms of split
octonions. The $SU\,(3)$ group appears as the automorphism group
of octonion representation leaving the complex subspace and the scalar
product invariant. This approach is extended by many \cite{key-14,key-15,key-16}
to investigate the role of octonions and division algebra in unified
gauge theories, higher dimensional theories of supersymmetry and super
strings. Octonions are also used by Buoncristiani \cite{key-17} in
writing Yang-Mill's field equation in a simpler form. The extension
of quaternion matrices to octonions for their interpretation in non-Riemannian
geometry is described by Marques et al in a series of papers \cite{key-18,key-19,key-20,key-21,key-22}. 

During the past few years, there has been considerable interest in
higher-dimensional kinematical models \cite{key-23,key-24} for a
proper and unified representation of relativistic object, bradyonic
as well as tachyonic (including those with internal structure). It
has been speculated \cite{key-25,key-26,key-27} that the problem
of representation and localization of tachyonic objects may be solved
only with the extension of four-dimensional Minkowski space to higher-dimensional
space-time. It has already been discussed in a series of papers \cite{key-28,key-29,key-30,key-31,key-32,key-33}
that the true localization space for the representation of tachyons
is $T^{4}-$ space with one space and three time coordinates while
that for bradyons is the usual $R^{4}-$ space with one time and three
space coordinates. The unified eight dimensional space time has been
discussed earlier \cite{key-32,key-33} as the unified space of bradyons
and tachyons i.e. $R^{8}=R^{4}\,\bigcup\, T^{4}$. The two $R^{4}-$
and $T^{4}-$localization spaces are considered as external (internal)
space for bradyons ( tachyons) and vice versa. Moreover, the built-in
duality associated with the combination of symmetries of $R^{4}-$
and $T^{4}-$ subspaces is useful to understand the problem of quark
confinement in quantum chromodynamics where the role of tricolors
would be played by three time coordinates for bradyons and those of
three space coordinates for tachyons.

In order to interpret octonion wave equations in eight dimensional
space- time, in the present paper, we have made an attempt to discuss
the octonion field equation as the equation of motion for particles
carrying simultaneously electric and magnetic (monopole \cite{key-34}
) charges (i.e. dyons \cite{key-35} ). Starting with the regularity
condition in the octonion field equation we have developed consistent
and compact formulation of eight-dimensional potential and current
equations of dyons. It has been demonstrated that the component of
octonion potential wave equations behaves neither the generalized
electromagnetic fields of monopoles nor the dyons. Rather, they have
the mixed behaviour of electromagnetic fields associated with the
electric and magnetic charges in external and internal spaces.

After decomposing the octonion wave equation in to two quaternion-valued
wave equations, it is shown that the two different spaces demonstrate
the separate wave equations for electromagnetic fields. Visualizing
the external four spaces as the localization space for bradyons and
the internal space as the localization space for tachyons, it is shown
that the octonion wave equation reduces to the Maxwell's equation
(field equation) for bradyons in $R^{4}$ - space as well as that
for tachyons in $T^{4}$- space. It has also been emphasized that
the quaternionic decomposition of octonion wave equation gives rise
to the equations of dyons and tachyonic dyons, when we take them as
combination of bi-quaternion instead of real quaternion in external
and internal four-dimensional spaces of eight dimensional space time.

Then we have made an attempt to investigate the split octonion wave
equation and its interpretation in classical electrodynamics. Split
octonion electrodynamics has been deduced in terms of Zorn's vector
matrix realization by describing electrodynamics potential, current
and other dynamical quantities as octonion variables. Also the consistent
and compact forms of eight-dimensional potential and current equation
of dyons are obtained in terms of Zorn's vector matrix realization
of split octonions. It has been shown that the octonion valued potential
wave equation behaves neither the generalized electromagnetic fields
of monopoles nor of the dyons. Rather, they have the mixed behaviour
of electromagnetic fields associated with the electric and magnetic
charges in external and internal spaces. At last, it is shown that
split octonion Zorn's vector realization reproduces two different
spaces to demonstrate the separate wave equations for electromagnetic
fields. Visualizing the external four space as the localization space
for tachyons, it is shown that the octonion wave equation, when expressed
in terms of split octonions, reduces to the Maxwell's equation (field
equation) for bradyons in $R^{4}$- space as well as that for tachyons
in $T^{4}$- space in the absence of other.

\section{OCTONION DEFINITION}

An octonion $x$ is expressed as a set of eight real numbers

\begin{eqnarray}
x & = & e_{0}x_{0}+e_{1}x_{1}+e_{2}x_{2}+e_{3}x_{3}+e_{4}x_{4}+e_{5}x_{5}+e_{6}x_{6}+e_{7}x_{7}=e_{0}x_{0}+\sum_{A=1}^{7}e_{A}x_{A}\label{eq:1}\end{eqnarray}
where $e_{A}(A=1,2,...,7)$ are imaginary octonion units and $e_{0}$
is the multiplicative unit element. Set of octets $(e_{0},e_{1},e_{2},e_{3},e_{4},e_{5},e_{6},e_{7})$
are known as the octonion basis elements and satisfy the following
multiplication rules

\begin{eqnarray}
e_{0}=1;\,\, & e_{0}e_{A}=e_{A}e_{0}=e_{A}; & \,\,\,\, e_{A}e_{B}=-\delta_{AB}e_{0}+f_{ABC}e_{C}.\,\,(A,B,C=1,2,.....,7)\label{eq:2}\end{eqnarray}
The structure constants $f_{ABC}$ is completely antisymmetric and
takes the value $1$ for following combinations,

\begin{eqnarray}
f_{ABC}= & +1; & \forall(ABC)=(123),\,(471)\,(257)\,(165)\,(624)\,(543)\,(736).\label{eq:3}\end{eqnarray}
It is to be noted that the summation convention is used for repeated
indices. Here the octonion algebra $\mathcal{O}$ is described over
the algebra of rational numbers having the vector space of dimension
$8$. We now get the following relations among octonion basis elements 

\begin{eqnarray}
\left[e_{A},\,\, e_{B}\right] & = & 2\, f_{ABC}\, e_{C};\nonumber \\
\left\{ e_{A},\,\, e_{B}\right\}  & = & -2\,\delta_{AB}e_{0};\nonumber \\
e_{A}(\, e_{B}\, e_{C}) & \neq & (e_{A}\, e_{B}\,)\, e_{C}\label{eq:4}\end{eqnarray}
where brackets $[\,\,]$ and $\{\,\,\}$ are used respectively for
commutation and the anti commutation relations while $\delta_{AB}$
is the usual Kronecker delta-Dirac symbol. Octonion conjugate is defined
as

\begin{eqnarray}
\overline{x} & = & e_{0}x_{0}-e_{1}x_{1}+-e_{2}x_{2}+-e_{3}x_{3}-e_{4}x_{4}-e_{5}x_{5}-e_{6}x_{6}-e_{7}x_{7}=e_{0}x_{0}-\sum_{A=1}^{7}e_{A}x_{A}\label{eq:5}\end{eqnarray}
where we have used the conjugates of basis elements as $\overline{e_{0}}=e_{0}$
and $\overline{e_{A}}=-e_{A}$. Hence an octonion can be decomposed
in terms of its scalar $(Sc(x))$ and vector $(Vec(x))$ parts as 

\begin{eqnarray}
Sc(x) & = & \frac{1}{2}(\, x\,+\,\overline{x}\,);\,\,\,\,\,\, Vec(x)=\frac{1}{2}(\, x\,-\,\overline{x}\,)=\sum_{A=1}^{7}\, e_{A}x_{A}.\label{eq:6}\end{eqnarray}
Conjugates of product of two octonions and its own are described as

\begin{eqnarray}
\overline{(x\, y)}= & \overline{y}\,\,\overline{x};\,\,\,\,\,\, & \overline{(\overline{x})}=x.\label{eq:7}\end{eqnarray}
while the scalar product of two octonions is defined as 

\begin{eqnarray}
\left\langle x\,,\, y\right\rangle  & =\frac{1}{2}(x\,\overline{y}+y\,\overline{x})=\frac{1}{2}(\overline{x}\, y+\overline{y}\, x)= & \sum_{\alpha=0}^{7}\, x_{\alpha}\, y_{\alpha}.\label{eq:8}\end{eqnarray}
The norm $N(x)$ and inverse $x^{-1}$(for a nonzero $x$) of an octonion
are respectively defined as

\begin{eqnarray}
N(x)=x\,\overline{x}=\overline{x}\, x & = & \sum_{\alpha=0}^{7}\, x_{\alpha}^{2}.e_{0};\nonumber \\
x^{-1} & = & \frac{\overline{x}}{N(x)}\,\Longrightarrow x\, x^{-1}=x^{-1}\, x=1.\label{eq:9}\end{eqnarray}
The norm $N(x)$ of an octonion $x$ is zero if $x=0$, and is always
positive otherwise. It also satisfies the following property of normed
algebra

\begin{eqnarray}
N(x\, y)= & N(x)\, N(y)= & N(y)\, N(x).\label{eq:10}\end{eqnarray}
Equation (\ref{eq:4}) shows that octonions are not associative in
nature and thus do not form the group in their usual form. Non-associativity
of octonion algebra $\mathcal{O}$ is provided by the associator $(x,y,z)=(xy)z-x(yz)\,\,\forall x,y,z\in\mathcal{O}$
defined for any $3$ octonions. If the associator is totally antisymmetric
for exchanges of any $2$ variables, i.e. $(x,y,z)=-(z,y,x)=-(y,x,z)=-(x,z,y)$,
then the algebra is called alternative. Hence, the octonion algebra
is neither commutative nor associative but, is alternative.

\section{OCTONION WAVE EQUATION}

In order to write the Octonion wave equation, let us define the differential
octonion $D$ as

\begin{eqnarray}
D & = & \sum_{\mu=0}^{7}e_{\mu}D_{\mu}\label{eq:11}\end{eqnarray}
where $D_{\mu}$ are described as the components of differential operator
in eight dimensional representation. Here we assume the eight - dimensional
space as the combination of two ( external and internal ) four-dimensional
spaces. As such, we describe a function of octonion variable as\begin{eqnarray}
\mathcal{F}(X) & =\sum_{\mu=0}^{7} & e_{\mu}f_{\mu}(X)=f_{0}+e_{1}f_{1}+e_{2}f_{2}+.....+e_{7}f_{7}\label{eq:12}\end{eqnarray}
where $f_{\mu}$ are scalar functions. Since octonions are neither
commutative nor associative, one has to be very careful to multiply
the octonion either from left or from right in terms of regularity
conditions. As such, a function $\mathcal{F}(X)$ of an octonion variable
$X={\displaystyle \sum_{\mu=0}^{7}e_{\mu}X_{\mu}}$ is left regular
at $X$ if and only if $\mathcal{F}(X)$ satisfies the condition

\begin{eqnarray}
D\mathcal{F}(X) & = & 0.\label{eq:13}\end{eqnarray}
Similarly, a function $G(X)$ is a right regular if and only if

\begin{eqnarray}
G(X)D & = & 0\label{eq:14}\end{eqnarray}
where $G(X)=g_{0}+g_{1}e_{1}+g_{2}e_{2}+.....+g_{7}e_{7}.$ Then we
get 

\begin{eqnarray}
D\mathcal{F} & = & I_{0}+I_{1}e_{1}+I_{2}e_{2}+I_{3}e_{3}+I_{4}e_{4}+I_{5}e_{5}+I_{6}e_{6}+I_{7}e_{7}\label{eq:15}\end{eqnarray}
where

\begin{eqnarray}
I_{0} & = & \partial_{0}f_{0}-\partial_{1}f_{1}-\partial_{2}f_{2}-\partial_{3}f_{3}-\partial_{4}f_{4}-\partial_{5}f_{5}-\partial_{6}f_{6}-\partial_{7}f_{7};\nonumber \\
I_{1} & = & \partial_{0}f_{1}+\partial_{1}f_{0}+\partial_{2}f_{3}-\partial_{3}f_{2}+\partial_{6}f_{5}-\partial_{5}f_{6}-\partial_{7}f_{4}+\partial_{4}f_{7};\nonumber \\
I_{2} & = & \partial_{0}f_{2}+\partial_{2}f_{0}+\partial_{3}f_{1}-\partial_{1}f_{3}+\partial_{4}f_{6}-\partial_{6}f_{4}-\partial_{7}f_{5}+\partial_{5}f_{7};\nonumber \\
I_{3} & = & \partial_{0}f_{3}+\partial_{3}f_{0}+\partial_{1}f_{2}-\partial_{2}f_{1}+\partial_{6}f_{7}-\partial_{7}f_{6}+\partial_{5}f_{4}-\partial_{4}f_{5};\nonumber \\
I_{4} & = & \partial_{0}f_{4}+\partial_{4}f_{0}+\partial_{3}f_{5}-\partial_{5}f_{3}-\partial_{2}f_{6}+\partial_{6}f_{2}-\partial_{1}f_{7}+\partial_{7}f_{1};\nonumber \\
I_{5} & = & \partial_{0}f_{5}+\partial_{5}f_{0}+\partial_{1}f_{6}-\partial_{6}f_{1}+\partial_{7}f_{2}-\partial_{2}f_{7}-\partial_{3}f_{4}+\partial_{4}f_{3};\nonumber \\
I_{6} & = & \partial_{0}f_{6}+\partial_{6}f_{0}-\partial_{1}f_{5}+\partial_{5}f_{1}+\partial_{2}f_{4}-\partial_{4}f_{2}-\partial_{3}f_{7}+\partial_{7}f_{3};\nonumber \\
I_{7} & = & \partial_{0}f_{7}+\partial_{7}f_{0}+\partial_{1}f_{4}-\partial_{4}f_{1}+\partial_{2}f_{5}-\partial_{5}f_{2}-\partial_{6}f_{3}+\partial_{3}f_{6}.\label{eq:16}\end{eqnarray}
The regularity condition (\ref{eq:13}) may now be considered as homogeneous
octonion wave equation for octonion variables without sources. On
the other hand, equation (\ref{eq:15}) is considered as the inhomogeneous
wave equation 

\begin{eqnarray}
D\mathcal{F} & = & I\label{eq:17}\end{eqnarray}
where $I$ is again an octonion. Similarly, we may also write the
homogeneous as well as inhomogeneous octonion wave equations on using
the right regularity condition (\ref{eq:14}). We may now interpret
these octonion wave equations as the classical wave (field) equations
of physical variables.

\subsection{POTENTIAL EQUATION FOR ELECTROMAGNETIC FIELDS WITH SOURCES}

Let us consider the case of generalized electromagnetic fields of
dyons (particles carrying simultaneous existence of electric and magnetic
charges). We may now define an octonion valued potential, in eight
dimensional formalism as the combinations of two four-dimensional
spaces, as follows

\begin{eqnarray}
\emptyset & =\sum_{\mu=0}^{7} & e_{\mu}\emptyset_{\mu}=\sum_{\mu=0}^{3}e_{\mu}A_{\mu}+\sum_{\nu=4}^{7}e_{\nu}B_{\nu};\label{eq:18}\end{eqnarray}
where $\emptyset$ is the octonion potential and $A_{\mu}$ and $B_{\nu}$are
assumed respectively as the electric and magnetic four - potentials
associated with the electric and magnetic charges of dyons \cite{key-36}.
As such, we may write the wave equation for octonion potential variable
for simultaneous existence of electric and magnetic charges on a particle
(namely dyons) in the following manner ,

\begin{eqnarray}
\overline{D}\emptyset & = & F\label{eq:19}\end{eqnarray}
where

\begin{eqnarray}
\overline{D} & = & F\sum_{\mu=0}^{7}\overline{e_{\mu}}\, D_{\mu}=e_{0}D_{0}-\sum_{A=1}^{7}e_{A}D_{A};\,\,\,\,\,\, F=\sum_{\mu=0}^{7}F_{\mu}e_{\mu}.\label{eq:20}\end{eqnarray}
The coefficients of octonion $F$ are then given by 

\begin{eqnarray}
F_{0} & = & \partial_{0}\emptyset_{0}+\partial_{1}\emptyset_{1}+\partial_{2}\emptyset_{2}+\partial_{3}\emptyset_{3}+\partial_{4}\emptyset_{4}+\partial_{5}\emptyset_{5}+\partial_{6}\emptyset_{6}+\partial_{7}\emptyset_{7};\nonumber \\
F_{1} & = & \partial_{0}\emptyset_{1}-\partial_{1}\emptyset_{0}+\partial_{3}\emptyset_{2}-\partial_{2}\emptyset_{3}+\partial_{7}\emptyset_{4}-\partial_{4}\emptyset_{7}+\partial_{5}\emptyset_{6}-\partial_{6}\emptyset_{5};\nonumber \\
F_{2} & = & \partial_{0}\emptyset_{2}-\partial_{2}\emptyset_{0}+\partial_{1}\emptyset_{3}-\partial_{3}\emptyset_{1}+\partial_{6}\emptyset_{4}-\partial_{4}\emptyset_{6}+\partial_{7}\emptyset_{5}-\partial_{5}\emptyset_{7};\nonumber \\
F_{3} & = & \partial_{0}\emptyset_{3}-\partial_{3}\emptyset_{0}+\partial_{2}\emptyset_{1}-\partial_{1}\emptyset_{2}+\partial_{4}\emptyset_{5}-\partial_{5}\emptyset_{4}+\partial_{7}\emptyset_{6}-\partial_{6}\emptyset_{7};\nonumber \\
F_{4} & = & \partial_{0}\emptyset_{4}-\partial_{4}\emptyset_{0}-\partial_{3}\emptyset_{5}+\partial_{5}\emptyset_{3}+\partial_{2}\emptyset_{6}-\partial_{6}\emptyset_{2}+\partial_{1}\emptyset_{7}-\partial_{7}\emptyset_{1};\nonumber \\
F_{5} & = & \partial_{0}\emptyset_{5}-\partial_{5}\emptyset_{0}-\partial_{1}\emptyset_{6}+\partial_{6}\emptyset_{1}-\partial_{7}\emptyset_{2}+\partial_{2}\emptyset_{7}+\partial_{3}\emptyset_{4}-\partial_{4}\emptyset_{3};\nonumber \\
F_{6} & = & \partial_{0}\emptyset_{6}-\partial_{6}\emptyset_{0}+\partial_{1}\emptyset_{5}-\partial_{5}\emptyset_{1}-\partial_{2}\emptyset_{4}+\partial_{4}\emptyset_{2}+\partial_{3}\emptyset_{7}-\partial_{7}\emptyset_{3};\nonumber \\
F_{7} & = & \partial_{0}\emptyset_{7}-\partial_{7}\emptyset_{0}-\partial_{1}\emptyset_{4}+\partial_{4}\emptyset_{1}-\partial_{2}\emptyset_{5}+\partial_{5}\emptyset_{2}+\partial_{6}\emptyset_{3}-\partial_{3}\emptyset_{6}.\label{eq:21}\end{eqnarray}
We get $F_{0}=0$ in the Euclidean space $(++++++)$ due to eight
dimensional Lorentz gauge condition (i.e the combination of two gauges
associated with electric and magnetic potentials). Thus, one dimensional
octonion representation is identical to eight dimensional spaces over
the field of real numbers. It is isomorphic to four-dimensional space
representation over the field of complex variables which is equivalent
to two-dimensional space representation over quaternion field variables.
Similarly, one dimensional quaternion space is isomorphic to four-dimensional
space over the field of real numbers which is identical to two- dimensional
space over the field of complex numbers. So, an octonionic potential
may also be described as the combination of two quaternion potentials
in the following manner;

\begin{eqnarray}
\emptyset & = & \emptyset_{a}+e_{7}\emptyset_{b};\,\,\,\, D=D_{a}+e_{7}D_{b};\,\,\,\, F=F_{a}+e_{7}F_{b}\label{eq:22}\end{eqnarray}
where $\emptyset_{a}$, $\emptyset_{b}$, $D_{a}$, $D_{b}$, $F_{a}$
and $F_{b}$ are quaternion variables described as

\begin{eqnarray}
\emptyset_{a} & = & \emptyset_{0}+e_{1}\emptyset_{1}+e_{2}\emptyset_{2}+e_{3}\emptyset_{3};\nonumber \\
\emptyset_{b} & = & \emptyset_{7}+e_{1}\emptyset_{4}+e_{2}\emptyset_{5}+e_{3}\emptyset_{6};\nonumber \\
D_{a} & = & \partial_{0}+e_{1}\partial_{1}+e_{2}\partial_{2}+e_{3}\partial_{3};\nonumber \\
D_{b} & = & \partial_{7}+e_{1}\partial_{4}+e_{2}\partial_{5}+e_{3}\partial_{6};\nonumber \\
F_{a} & = & F_{0}+e_{1}F_{1}+e_{2}F_{2}+e_{3}F_{3};\nonumber \\
F_{b} & = & F_{7}+e_{1}F_{4}+e_{2}F_{5}+e_{3}F_{6}.\label{eq:23}\end{eqnarray}
Here $e_{1}$, $e_{2}$, and $e_{3}$ ; three quaternion units; satisfy
the multiplication rule $e_{j}e_{k}=-\delta_{jk}+\varepsilon_{jkl}e_{l}\quad\forall(j,k,l=1,2,3)$.
If we replace the octonion unit $e_{7}$ by imaginary unit $i=\sqrt{-1}$,
commutating with all the octonion basis elements $e_{A}$, the resultant
theory of octonions become the theory of bi-quaternion variables where
generalized fields of dyons are written in compact, covariant and
consistent way \cite{key-37}. We may now consider the following mapping
for different four-dimensional spaces,

\begin{eqnarray}
\partial_{7}\rightarrow\partial_{0}^{'}, & \emptyset_{7}\rightarrow B_{0}, & \emptyset_{0}\rightarrow A_{0};\nonumber \\
\partial_{4}\rightarrow\partial_{1}^{'}, & \emptyset_{4}\rightarrow B_{1}, & \emptyset_{1}\rightarrow A_{1};\nonumber \\
\partial_{5}\rightarrow\partial_{2}^{'}, & \emptyset_{5}\rightarrow B_{2}, & \emptyset_{2}\rightarrow A_{2};\nonumber \\
\partial_{6}\rightarrow\partial_{3}^{'}, & \emptyset_{6}\rightarrow B_{3}, & \emptyset_{3}\rightarrow A_{3}.\label{eq:24}\end{eqnarray}
Then we get

\begin{eqnarray}
F_{j} & = & \partial_{0}A_{j}-\partial_{j}A_{0}+(\overrightarrow{\nabla}\times\overrightarrow{A})_{j}+\partial_{0}^{'}B_{j}-\partial_{j}^{'}B_{0}+(\overrightarrow{\nabla}\times\overrightarrow{B})_{j};\nonumber \\
F_{j+3} & = & \partial_{0}B_{j}-\partial_{j}B_{0}+(\overrightarrow{\nabla}\times\overrightarrow{B})_{j}+\partial_{0}^{'}A_{j}-\partial_{j}^{'}A_{0}+(\overrightarrow{\nabla}'\times\overrightarrow{A})_{j};\nonumber \\
F_{7} & = & \partial_{0}B_{0}-\partial_{1}B_{1}-\partial_{2}B_{2}-\partial_{3}B_{3}-(\partial_{0}^{'}A_{0}+\partial_{1}^{'}A_{1}+\partial_{2}^{'}A_{2}+\partial_{3}^{'}A_{3}).\label{eq:25}\end{eqnarray}
Hence equation (\ref{eq:25}) shows that $F_{0},\,\, F_{j},\,\, F_{j+3}$
and $F_{7}$ correspond to the components of electric and magnetic
fields obtained in terms of components of two four-potentials ( $A_{\mu}$
and $B_{\mu}$) in internal and external spaces. If these two spaces
are completely disjoint spaces, we get only $F_{j}$ ; and $F_{0}$
vanishing due to Lorentz gauge condition ; while $F_{j+3}\,$and $F_{7}$
do not occur. Here prime derivatives are associated with the second
four-dimensional structure of eight potential. In equation (\ref{eq:25}),
we find that $F_{j}$ is made up of the $j^{th}$ components of electric
and magnetic fields. Here the first term $\partial_{0}A_{j}-\partial_{j}A_{0}$
describes the electric field and second term $\overrightarrow{(\nabla}\times\overrightarrow{A})_{j}$corresponds
to the magnetic field in four-dimensional ( we call it as the external
) space. In the light of the duality mechanism between electric and
magnetic fields (and for electric $\left\{ A_{\mu}\right\} $ and
magnetic $\left\{ B_{\mu}\right\} $ four-potentials), the third term
$\partial_{0}B_{j}-\partial_{j}B_{0}$ of $F_{j}$ is equivalent to
the magnetic field while the last term $(\overrightarrow{\nabla}\times\overrightarrow{B})_{j}$
denotes the electric field in other four-dimensional ( let us call
it as internal or magnetic) space. On mixing of these two spaces,
$\partial_{0}A_{j}-\partial_{j}A_{0}+(\overrightarrow{\nabla}\times\overrightarrow{A})_{j}$
resembles with the expression of generalized electric field and $\partial_{0}^{'}B_{j}-\partial_{j}^{'}B_{0}+(\overrightarrow{\nabla}\times\overrightarrow{B})_{j}$
as that of the generalized magnetic fields of dyons \cite{key-36,key-37}
. So, it looks awkward to combine external (internal) space of electric
charge (magnetic monopole) to the internal (external) space of magnetic
monopole (electric charge) to interpret $F_{j}$ . As such, we may
conclude that the equation (\ref{eq:25}) does not represent the true
generalization of potential equation for the generalized electromagnetic
fields of dyons. Rather, we have obtained the mixed behaviour of electric
and magnetic charges in internal and external spaces. So, octonion
wave equation of electromagnetic potential faces difficulties and
hence needs modification to interpret it consistently.

\subsection{CURRENT EQUATION}

We now Analise the octonion wave equation (\ref{eq:17}) as the current
equation in eight dimensional representations i.e.

\begin{eqnarray}
D & F= & S\label{eq:26}\end{eqnarray}
where

\begin{eqnarray}
S_{0} & = & \partial_{0}F_{0}-\partial_{1}F_{1}-\partial_{2}F_{2}-\partial_{3}F_{3}-(\partial_{1}^{'}F_{4}+\partial_{2}^{'}F_{5}+\partial_{3}^{'}F_{6}+\partial_{0}^{'}F_{7});\nonumber \\
S_{j} & = & \partial_{0}F_{j}+\partial_{j}F_{0}+(\overrightarrow{\nabla}\times\overrightarrow{F})_{j}-\partial_{0}^{'}F_{j+3}+\partial_{j}^{'}F_{7}-(\overrightarrow{\nabla}'\times\overrightarrow{F})_{j+3};\nonumber \\
S_{j+3} & = & \partial_{0}^{'}F_{j}+\partial_{j}^{'}F_{0}-(\overrightarrow{\nabla}'\times\overrightarrow{F})_{j}-\partial_{0}F_{j+3}-\partial_{j}F_{7}-(\overrightarrow{\nabla}\times\overrightarrow{F})_{j+3};\nonumber \\
S_{7} & = & \partial_{1}F_{4}+\partial_{2}F_{5}+\partial_{3}F_{6}+\partial_{0}F_{7}+(\partial_{0}^{'}F_{0}-\partial_{1}^{'}F_{1}-\partial_{2}^{'}F_{2}-\partial_{3}^{'}F_{3}).\label{eq:27}\end{eqnarray}
Thus, the current equation for octonion variables describes \cite{key-38}
the generalized structure of differential equations (Maxwell's equations)
in eight - dimensional space-time. Four- dimensional reduction of
these differential equations may be visualized as the Maxwell's equations
in internal and external spaces. So, it is difficult to explain the
various terms associated with the eight parameters $S_{\Lambda}(\Lambda=0,1,2,3...,7)$correctly.
Silagade \cite{key-38} provided the interpretation of the homogeneous
octonion wave equation $DF=0$ as equivalent to one of the pair of
seven-dimensional Maxwell's equations and the second pair of seven
dimensional Maxwell's equations may be obtained on applying the duality
transformations between electric and magnetic fields therein. Similarly,
Gamba and Gogberashvili \cite{key-8} also tried to explain the structure
of eight parameters $S_{\Lambda}(\Lambda=0,1,2,3...,7)$but the consistent
justification needs modifications. Hence, octonion wave equation (\ref{eq:26})
may be identified rather a current equation or octonion field equation
in eight dimensional spaces for mixed structural behaviour of sources
(the electric and magnetic) instead of dyons. In other words we may
say that the equation (\ref{eq:26}) can not be described as the true
field equation of dyons. Here $S_{0}$ and $S_{7}$ may be taken vanishing
due to equations of continuity in internal and external spaces.

\subsection{QUATERNION DECOMPOSITION}

We may decompose an octonion in terms of two-quaternions. So, let
us decompose octonion wave equation for potential in terms of two
quaternions as

\begin{eqnarray}
\overline{D}\emptyset=(\overline{D}_{a}-e_{7}D_{b})(\emptyset_{a}+e_{7}\emptyset_{b}) & = & (\phi+e_{7}\varphi)=F.\label{eq:28}\end{eqnarray}
Here

\begin{eqnarray}
\phi & = & \overline{D}_{a}\emptyset_{a}+\emptyset_{b}\widetilde{D}_{b}=(D_{0}-D_{1}e_{1}-D_{2}e_{2}-D_{3}e_{3})\emptyset_{a}+\emptyset_{b}(\overleftarrow{D_{7}}-\overleftarrow{D_{4}}e_{1}-\overleftarrow{D_{5}}e_{2}-\overleftarrow{D_{6}}e_{3})\label{eq:29}\end{eqnarray}
where $[\overleftarrow{D_{r}},(r=4,5,6,7)]$ represents the partial
differential of $\emptyset_{b}$ from right to left , $\widetilde{D}_{b}$
denotes the quaternion conjugation $(\tilde{e_{o}}=e_{0};\quad\tilde{e_{j}}=-e_{j}\quad\forall j=1,2,3)$
from right to left i.e. 

\begin{eqnarray*}
\overline{D}_{a} & = & D_{0}-D_{1}e_{1}-D_{2}e_{2}-D_{3}e_{3};\end{eqnarray*}

\begin{eqnarray*}
\overline{\widetilde{D}_{a}} & = & D_{0}+D_{1}e_{1}+D_{2}e_{2}+D_{3}e_{3}=D_{1};\end{eqnarray*}

\begin{eqnarray*}
\widetilde{D}_{b} & = & D_{7}-D_{4}e_{1}-D_{5}e_{2}-D_{6}e_{3}\end{eqnarray*}
and

\begin{eqnarray}
\varphi & =\overline{\widetilde{D}}_{a} & \emptyset_{b}-\emptyset_{a}D_{b}=D_{a}\emptyset_{b}-\emptyset_{a}D_{b}.\label{eq:30}\end{eqnarray}
Here an octonion-valued potential is described as the sum of two quaternions
by Cayley-Dickson doubling process. Thus, we observe that the octonion
potential wave equation rules out the existence of isolated of magnetic
monopole and provides a mixed structural behaviour of electromagnetic
fields containing both charges simultaneously. In other words,  we
may say that it is a generalized field equation of electromagnetic
fields associated with a particle carrying simultaneous existence
of electric and magnetic charges (namely dyons). The generalized electromagnetic
fields of dyons are symmetric and dual invariant in terms of two four
potentials. So, the possibility of monopoles (or dyons) in non Abelian
or supersymmetric gauge theories is directly linked with the existence
of second four potential which is hidden in external space and present
in internal space while the electric four potential is the consequence
of our external space and seems to be hidden in internal space.

Similarly we may express the current equation given by equation (\ref{eq:26})
in terms of two quaternions as

\begin{eqnarray}
DF & =(D_{a}+e_{7}D_{b})(g+e_{7}h) & =S=s_{a}+e_{7}s_{b}\label{eq:31}\end{eqnarray}
where

\begin{eqnarray*}
s_{a} & = & D_{a}g-hD_{b}=D_{a}\overline{D}_{b}\emptyset_{a}+\emptyset_{b}D_{b}\overline{D}_{b}\end{eqnarray*}

\begin{eqnarray}
s_{b} & = & \widetilde{D}_{a}h+gD_{b}=\overline{D}_{a}D_{a}\emptyset_{b}+\emptyset_{b}\overline{D_{b}}D_{b}.\label{eq:32}\end{eqnarray}
Thus $S=s_{a}+e_{7}s_{b}$ is the octonion expression for generalized
current density. Similarly, we may express the generalized force and
generalized field tensor density in terms of two quaternions.

\subsection{LOCALIZATION SPACES OF BRADYONS AND TACHYONS}

~~~~~ We have described the octonion (eight dimensional) space
is made of two quaternion (namely the external and internal four dimensional)
spaces. Let us suppose the external four space as the usual Minikowski
(or Euclidean) $R^{4}\Rightarrow M\,(1,3)$ - space (consisting one
time and three space coordinates). This space has been named \cite{key-23,key-24}
as. the localization space of bradyons (particles traveling slower
than light$\rightarrow$subluminal particles). Accordingly, we describe
the internal four dimensional space as the $T^{4}\Rightarrow M\,(3,1)$
- space (consisting three time and one space coordinates). The possibility
of such space is explored \cite{key-23,key-24} as the localization
space of tachyons (particles traveling faster than light$\rightarrow$superluminal
particles). Hence an octonion eight dimensional space is described
as the unified space containing both external $R^{4}\Rightarrow M\,(1,3)$
and internal $T^{4}\Rightarrow M\,(3,1)$ four - dimensional spaces.
In other words we may identify the octonion space as the unified localization
space for the description of bradyons and tachyons. It is because
that the bradyons are the objects of our four dimensional world for
which the Cauchy's data are described on the plane $\left\{ t=0\right\} $while
tachyons are localized in time and their Cauchy's data lies on the
plane $\left\{ x=0\right\} .$ Hence, the eight dimensional space
of octonion variables described above is the unified localization
space for bradyons and tachyons and hence be expressed as $R^{8}=R^{4}\,\bigcup\, T^{4}$.
The $T^{4}-$space is characterized as the hidden space for bradyons
while $R^{4}-$space is identified as the hidden space for tachyons.
We may also interpret that the $R^{4}-$ space is visualized as the
internal space for tachyons (and external space for bradyons) and
correspondingly the $T^{4}-$ space is the internal space for bradyons
(and external space for tachyons). First quaternion variable of equation
(\ref{eq:31}) maps to four-dimensional space for bradyons i.e. $R^{4}\simeq(t,\overrightarrow{r})$
while the second quaternion variable gives rise to four-dimensional
space for tachyons i.e. $T^{4}\simeq(r,\overrightarrow{t})$. In the
absence of second quaternion, equation (\ref{eq:32}) reduces to

\begin{eqnarray}
s_{a}= & D_{a}\overline{D}_{a}\emptyset_{a} & \Rightarrow(\frac{\partial^{2}}{\partial t^{2}}-\frac{\partial^{2}}{\partial x^{2}}-\frac{\partial^{2}}{\partial y^{2}}-\frac{\partial^{2}}{\partial z^{2}})\phi=\square(R)\phi\label{eq:33}\end{eqnarray}
which resembles to Cauchy-Feuter equation of quaternion variables
in $R^{4}\simeq(t,\overrightarrow{r})$ space and thus describes the
usual Maxwell's equations in electromagnetic fields in $R^{4}-$ space.
On the other hand in absence of first quaternion, equation (\ref{eq:32})
reduces to 

\begin{eqnarray}
s_{b}=\emptyset_{b} & \overline{D}_{b}D_{b} & \Rightarrow\varphi(\frac{\partial^{2}}{\partial r^{2}}-\frac{\partial^{2}}{\partial t_{x}^{2}}-\frac{\partial^{2}}{\partial t_{y}^{2}}-\frac{\partial^{2}}{\partial t_{z}^{2}})=\varphi\overleftarrow{\square}(T)\label{eq:34}\end{eqnarray}
which also describes the Cauchy - Feuter equation of quaternion variables
in $T^{4}\simeq(r,\overrightarrow{t})$ space and coincides with the
Maxwell's equations of superluminal photons in $T^{4}$ - space. Equation
(\ref{eq:33}) is the left regular while equation (\ref{eq:34}) is
the right regular. hence, the regularity conditions are changes for
internal and external spaces. In case of bi-quaternion, equations
(\ref{eq:33}) and (\ref{eq:34}) characterize respectively to the
equations of dyons for subluminal and superluminal electromagnetic
fields. In general (i.e. octonionic space) describes the unified structure
of subluminal and superluminal objects. As such, the octonionic eight
dimensional representation is the unified picture of bradyons and
tachyons and the octonionic current equations thus reproduce two different
kinds of Maxwell's equations in external and internal spaces in order
to explain the simultaneous description of bradyons and tachyons.
Here, we have developed the octonion wave equation and theory of octonion
field variables in compact and simpler manner. So, in order to overcome
the non-associativity, it is necessary to decompose them in terms
of two quaternions isomorphic to two different four-dimensional spaces.
The compact and simpler form of octonion field equation provide an
unified model of the theories of subluminal and superluminal electromagnetic
fields in view of the localizability of bradyons and tachyons in consistent
manner.

\section{SPLIT OCTONION WAVE EQUATION }

The split octonions are a non associative extension of quaternions
(or the split quaternions).They differ from the octonion in the signature
of quadratic form. The split octonions have a signature $(4,4)$ whereas
the octonions have positive signature $(8,0)$. The Cayley algebra
of octonions over the field of complex numbers $\mathbb{C_{C}=\mathbb{C\otimes C}}$
is visualized as the algebra of split octonions with basis elements
$u_{0}=\frac{1}{2}(e_{0}+ie_{7});\quad u_{0}^{\star}=\frac{1}{2}(e_{0}-ie_{7}),\quad u_{j}=\frac{1}{2}(e_{j}+ie_{j+3}),\quad u_{j}^{\star}=\frac{1}{2}(e_{j}-ie_{j+3})\,(j=1,2,3;i=\sqrt{-1})$
as the bi-valued (or bi - dimensional) representations of quaternion
units $e_{0},\quad e_{1}\quad,e_{2}\quad,e_{3}$ . The split Cayley
(octonion) algebra is thus expressed in terms of $2\times2$ Zorn's
vector matrices components of which are scalar and vector parts of
a quaternion. As such , we may also write an arbitrary split octonion
$A$ in terms of following $2\times2$ Zorn's vector matrix realizations
\cite{key-39} as

\begin{eqnarray}
A & =au_{0}^{\star}+bu_{0}+x_{i}u_{i}^{\star}+y_{i}u_{i} & =\left(\begin{array}{cc}
a & -\overrightarrow{x}\\
\overrightarrow{y} & b\end{array}\right).\label{eq:35}\end{eqnarray}
Split octonion conjugation of equation (\ref{eq:35}) is then described
as,

\begin{eqnarray}
\overline{A} & =au_{0}+bu_{0}*-x_{i}u_{i}*-y_{i}u_{i} & =\left(\begin{array}{cc}
b & \overrightarrow{x}\\
-\overrightarrow{y} & a\end{array}\right).\label{eq:36}\end{eqnarray}
The norm of $A$ is defined as,

\begin{eqnarray}
\overline{A}A & =A\overline{A} & =(ab+\overrightarrow{x}.\overrightarrow{y})\hat{1}\label{eq:37}\end{eqnarray}
where $\hat{1}$ is the identity element given by $\hat{1}=1u_{0}+1u_{0}^{\star}$.
Any four - vector $A_{\mu}$ (complex or real) can equivalently be
written in the following Zorn's matrix realization as 

\begin{eqnarray}
Z(A) & = & \left(\begin{array}{cc}
x_{4} & -\overrightarrow{x}\\
\overrightarrow{y} & y_{4}\end{array}\right)\label{eq:38}\end{eqnarray}
and

\begin{eqnarray}
Z(\overline{A}) & = & \left(\begin{array}{cc}
x_{4} & \overrightarrow{x}\\
-\overrightarrow{y} & y_{4}\end{array}\right).\label{eq:39}\end{eqnarray}
In equation (\ref{eq:38}, \ref{eq:39}), putting $\overrightarrow{x}=\overrightarrow{y}$,
we get the equivalent matrix realization for a four vector in bi-valued
four-dimensional Euclidean space - time. In order to develop the the
eight dimensional split octonion kinematics, we start with the following
definition of the split octonion differential operator $D$ and its
conjugate $\overline{D}$ in terms of $2\times2$ Zorn's matrix realizations,

\begin{eqnarray}
D & = & \left(\begin{array}{cc}
\partial_{4} & -\overrightarrow{\nabla}\\
\overrightarrow{\nabla} & \partial'_{4}\end{array}\right)\label{eq:40}\end{eqnarray}

\begin{eqnarray}
\overline{D} & = & \left(\begin{array}{cc}
\partial_{4} & \overrightarrow{\nabla}\\
-\overrightarrow{\nabla} & \partial'_{4}\end{array}\right)\label{eq:41}\end{eqnarray}
with the assumption that the primed variables are represented in internal
space whereas the unprimed variables are defined in external four
- dimensional space.

\subsection{POTENTIAL EQUATION}

Let us express an octonion potential in the following form \begin{eqnarray}
\emptyset & = & A_{4}u_{0}*+B_{4}u_{0}+B_{i}u_{i}*+A_{i}u_{i}\label{eq:42}\end{eqnarray}
where $\left\{ A_{\mu}\right\} =(A_{4},A_{j})$ and $\left\{ B_{\mu}\right\} =(B_{4},B_{j})$
are two four - potentials respectively associated with electric and
magnetic charges. Equation (\ref{eq:42}) may then be expressed in
terms of split octonion $2\times2$ Zorn's vector matrix realization
as

\begin{eqnarray}
\emptyset & = & \left(\begin{array}{cc}
A_{4} & -\overrightarrow{B}\\
\overrightarrow{A} & B_{4}\end{array}\right)\label{eq:43}\end{eqnarray}
where $A_{4}=i\,\emptyset_{e}$($e$ denotes the electric charge)
and $B_{4}=i\,\emptyset_{g}$($g$ denotes the magnetic charge). Here
$\overrightarrow{A}$ and $A_{4}$ are assumed to be the components
of electric four - potential $A_{\mu}=(\overrightarrow{A},\, i\,\emptyset_{e}).$
Similarly $\overrightarrow{B}$ and $B_{4}$ are considered as the
components of magnetic four-potential $B_{\mu}=(\overrightarrow{B},\, i\,\emptyset_{g})$.
We have designed the eight dimensional space spanned by split octonion
basis elements in terms of two four - dimensional space (namely the
external and internal space). Here also the electric four-potential
is described in external four-dimensional space while the magnetic
four - potential has been considered in internal four dimensional
spaces. Since electric charge and magnetic monopoles are dual to each
other ,we may interpret these two spaces as dual to each other. The
internal space of electric charge may be identified as the external
space of monopole while the external space of electric charge will
be visualized as the internal space of magnetic monopole or vice versa.
Accordingly, the electric charge is considered in external space and
magnetic charge in internal space. Hence, eight dimensional spaces
have the built in duality where a bradyonic monopole plays the role
of tachyonic electric charge while the tachyonic monopole looks like
a bradyonic electric charge or vice versa. Eight dimensional spaces
have been considered as the unification of $R^{4}-$ (bradyonic) and
$T^{4}-$ (tachyonic) spaces. 

Let us now write the the inhomogeneous octonion wave equation in its
split form as

\begin{eqnarray}
\overline{D}\emptyset & = & \left(\begin{array}{cc}
\partial_{4}A_{4}+\overrightarrow{\nabla}.\overrightarrow{A} & -(\partial_{4}\overrightarrow{B}-\overrightarrow{\nabla}B_{4}-\overrightarrow{\nabla}\times\overrightarrow{A})\\
\partial'_{4}\overrightarrow{A}-\overrightarrow{\nabla}'A_{4}-\overrightarrow{\nabla}'\times\overrightarrow{B} & \partial'_{4}B_{4}+\overrightarrow{\nabla}'.\overrightarrow{B}\end{array}\right)=F\label{eq:44}\end{eqnarray}
where $F$ is a field like electromagnetic field in split octonion
$2\times2$ Zorn's vector matrix realization as

\begin{eqnarray}
F & = & \left(\begin{array}{cc}
f_{4} & -\overrightarrow{f}\\
\overrightarrow{f}\,' & f_{4}'\end{array}\right).\label{eq:45}\end{eqnarray}
Comparing equations (\ref{eq:44}) and (\ref{eq:45}), we get

\begin{eqnarray}
f_{4} & = & \partial_{4}A_{4}+\overrightarrow{\nabla}.\overrightarrow{A}=0;\label{eq:46}\end{eqnarray}

\begin{eqnarray}
f_{4}' & = & \partial'_{4}B_{4}+\overrightarrow{\nabla}'.\overrightarrow{B}=0;\label{eq:47}\end{eqnarray}

\begin{eqnarray}
\overrightarrow{f} & = & \partial_{4}\overrightarrow{B}-\overrightarrow{\nabla}B_{4}-\overrightarrow{\nabla}\times\overrightarrow{A};\label{eq:48}\end{eqnarray}

\begin{eqnarray}
\overrightarrow{f}' & = & \partial'_{4}\overrightarrow{A}-\overrightarrow{\nabla}'A_{4}-\overrightarrow{\nabla}'\times\overrightarrow{B}.\label{eq:49}\end{eqnarray}
In equations (\ref{eq:46}, \ref{eq:47}), we could get $f_{4}=0$
and $f_{4}'=0$ on applying Lorentz gauge conditions on electric and
magnetic four potentials. We may now discuss the different cases in
the following sub sections:

\subsubsection*{(i) BRADYONIC CASE}

We may describe the theory of bradyons by substituting $B_{4}=A_{4},\,\overrightarrow{B}\,=\overrightarrow{A}$
in equations (\ref{eq:40}-\ref{eq:49}) and correspondingly $\partial_{4}=\partial_{4}',\,\overrightarrow{\nabla}=\overrightarrow{\nabla}'=\partial$.
Then we get

\begin{eqnarray}
\overline{D}\emptyset & =F & =\left(\begin{array}{cc}
\partial_{4}A_{4}+\overrightarrow{\nabla}.\overrightarrow{A} & -(\partial_{4}\overrightarrow{A}-\overrightarrow{\nabla}A_{4}-\overrightarrow{\nabla}\times\overrightarrow{A})\\
\partial_{4}\overrightarrow{A}-\overrightarrow{\nabla}A_{4}-\overrightarrow{\nabla}\times\overrightarrow{A} & \partial_{4}A_{4}+\overrightarrow{\nabla}.\overrightarrow{A}\end{array}\right).\label{eq:50}\end{eqnarray}
Using the definition of electric and magnetic fields given by

\begin{eqnarray}
\overrightarrow{E} & =-\frac{\partial\overrightarrow{A}}{\partial t}-\overrightarrow{\nabla}\emptyset_{e} & =-\frac{\partial\overrightarrow{A}}{\partial t}-\overrightarrow{\nabla}\phi_{e};\label{eq:51}\end{eqnarray}

\begin{eqnarray}
\overrightarrow{\mathcal{H}} & = & \overrightarrow{\nabla}\times\overrightarrow{A}\label{eq:52}\end{eqnarray}
and imposing Lorentz-gauge condition $\partial_{4}A_{4}+\overrightarrow{\nabla}.\overrightarrow{A}=\frac{\partial\emptyset_{e}}{\partial t}+\overrightarrow{\nabla}.\overrightarrow{A}=0,\,$
and using $\overrightarrow{\psi}=\overrightarrow{E}-i\overrightarrow{\mathcal{H}}(i=\sqrt{-1})$
, we get

\begin{eqnarray}
-\partial_{4}\overrightarrow{A}+\overrightarrow{\nabla}A_{4}+\overrightarrow{\nabla}\times\overrightarrow{A} & = & -i\overrightarrow{E}+\overrightarrow{\mathcal{H}}=-i\overrightarrow{\psi}*;\label{eq:53}\end{eqnarray}

\begin{eqnarray}
\partial_{4}\overrightarrow{A}-\overrightarrow{\nabla}A_{4}-\overrightarrow{\nabla}\times\overrightarrow{A} & = & i\overrightarrow{E}-\overrightarrow{\mathcal{H}}=i\overrightarrow{\psi}*.\label{eq:54}\end{eqnarray}
Thus equation (\ref{eq:50}) reduces to

\begin{eqnarray}
\overline{D}\emptyset & = & F\label{eq:55}\end{eqnarray}
where

\begin{eqnarray}
F & = & \left(\begin{array}{cc}
0 & -\overrightarrow{F}\\
\overrightarrow{F} & 0\end{array}\right)=i\psi*.\label{eq:56}\end{eqnarray}
Equation (\ref{eq:56}) is the split octonion form of field tensor
$F_{\mu\nu}=\partial_{\mu}A_{\nu}-\partial_{\nu}A_{\mu}$ describing
the electric and magnetic fields. Off diagonal vector components of
split octonions (\ref{eq:50}) are the conjugates to each other and
the scalar components along principal diagonal are associated with
the quaternion scalars (i.e. real quaternion). Similarly, we get

\begin{eqnarray}
D\overline{\emptyset} & = & F^{T}\label{eq:57}\end{eqnarray}
where

\begin{eqnarray}
F^{T} & = & \left(\begin{array}{cc}
0 & \overrightarrow{F}*\\
-\overrightarrow{F}* & 0\end{array}\right)=i\psi.\label{eq:58}\end{eqnarray}
and equations (\ref{eq:55}) and (\ref{eq:57}) lead to 

\begin{eqnarray}
\frac{1}{2}(\overline{D}\emptyset+D\overline{\emptyset}) & = & \left(\begin{array}{cc}
0 & \overrightarrow{\mathcal{H}}\\
-\overrightarrow{\mathcal{H}} & 0\end{array}\right)=\overrightarrow{\mathcal{H}}\label{eq:59}\end{eqnarray}

\begin{eqnarray}
\frac{1}{2}i\,(\overline{D}\emptyset-D\overline{\emptyset}) & = & \left(\begin{array}{cc}
0 & \overrightarrow{E}\\
-\overrightarrow{E} & 0\end{array}\right)=\overrightarrow{E}\label{eq:60}\end{eqnarray}
where $\overrightarrow{E}$ and $\overrightarrow{\mathcal{H}}$ are
the electric and magnetic fields given by equations (\ref{eq:51},\ref{eq:52}).
Here we have the different split octonion representation in comparison
to the case quaternion electrodynamics, where $\frac{1}{2}(\overline{D}\emptyset+D\overline{\emptyset})$
represents Lorentz condition (i.e. scalar part of quaternion) and
$\frac{1}{2}i\,(\overline{D}\emptyset-D\overline{\emptyset})$ represents
the vector part (pure quaternion) of a quaternion. It is obvious because
the off diagonal elements of an octonion is a quaternion while those
along principal diagonal are the scalars.

\subsubsection*{(ii) TACHYONIC CASE}

For the description of tachyons we substitute $A_{4}=B_{4},\,\overrightarrow{A}\,=\overrightarrow{B}$
in equations (\ref{eq:40}) to (\ref{eq:49}) and get the representation
of $T^{4}-$ space. In this space (one space and three time dimensions),
the scalar component is $r\,=|\overrightarrow{r}|=(x^{2}+y^{2}+z^{2})^{\frac{1}{2}}$and
the time gradient vector $\overrightarrow{\nabla}'=[t_{x},t_{y},t_{z},\partial_{4}'=(-ir)]$.
Hence the split octonion form of four differential operator, its conjugate
and the four potential of tachyons may be written as 

\begin{eqnarray}
D & = & \left(\begin{array}{cc}
\partial'_{4} & -\overrightarrow{\nabla}'\\
\overrightarrow{\nabla}' & \partial'_{4}\end{array}\right)\label{eq:61}\end{eqnarray}

\begin{eqnarray}
\overline{D} & = & \left(\begin{array}{cc}
\partial'_{4} & \overrightarrow{\nabla}'\\
-\overrightarrow{\nabla}' & \partial'_{4}\end{array}\right)\label{eq:62}\end{eqnarray}

\begin{eqnarray}
\emptyset & = & \left(\begin{array}{cc}
B_{4} & -\overrightarrow{B}\\
\overrightarrow{B} & B_{4}\end{array}\right).\label{eq:63}\end{eqnarray}
Thus we get 

\begin{eqnarray}
\overline{D}\emptyset & =F\,' & =\left(\begin{array}{cc}
\partial'_{4}B_{4}+\overrightarrow{\nabla}.\overrightarrow{B} & -(\partial'_{4}\overrightarrow{B}-\overrightarrow{\nabla}'B_{4}-\overrightarrow{\nabla}'\times\overrightarrow{B})\\
\partial'_{4}\overrightarrow{B}-\overrightarrow{\nabla}B_{4}-\overrightarrow{\nabla}'\times\overrightarrow{B} & \partial'_{4}B_{4}+\overrightarrow{\nabla}'.\overrightarrow{B}\end{array}\right).\label{eq:64}\end{eqnarray}
Now using the following expressions for tachyonic \cite{key-23,key-24}
electric and magnetic fields in $T^{4}$- space as

\begin{eqnarray}
\overrightarrow{E_{t}} & =-\frac{\partial\overrightarrow{B}}{\partial r}-\overrightarrow{\nabla}'\emptyset_{g} & =-\frac{\partial\overrightarrow{B}}{\partial r}-\overrightarrow{\nabla}'B_{4};\label{eq:65}\end{eqnarray}

\begin{eqnarray}
\overrightarrow{H_{t}} & = & \overrightarrow{\nabla}'\times\overrightarrow{B}\label{eq:66}\end{eqnarray}
and imposing Lorentz gauge condition in $T^{4}-$space as 

\begin{eqnarray*}
\partial'_{4}B_{4}+\overrightarrow{\nabla}'.\overrightarrow{B} & = & 0,\end{eqnarray*}
we get

\begin{eqnarray}
-\partial'_{4}\overrightarrow{B}+\overrightarrow{\nabla}'B_{4}+\overrightarrow{\nabla}'\times\overrightarrow{B} & = & -i\overrightarrow{E_{t}}+\overrightarrow{H_{t}}=-i\overrightarrow{\psi_{t}}*;\label{eq:67}\end{eqnarray}

\begin{eqnarray}
\partial'_{4}\overrightarrow{B}-\overrightarrow{\nabla}B_{4}-\overrightarrow{\nabla}'\times\overrightarrow{B} & = & i\overrightarrow{E_{t}}-\overrightarrow{H_{t}}=i\overrightarrow{\psi_{t}}*\label{eq:68}\end{eqnarray}
where $\overrightarrow{\psi_{t}}=\overrightarrow{E_{t}}-i\overrightarrow{H_{t}}$
($t$ denotes for tachyonic representations). Then equation (\ref{eq:64})
reduces to

\begin{eqnarray}
F\,' & = & \left(\begin{array}{cc}
0 & -\overrightarrow{F_{t}}\\
\overrightarrow{F_{t}} & 0\end{array}\right)=i\overrightarrow{\psi_{t}}*.\label{eq:69}\end{eqnarray}
Equation (\ref{eq:64}) is the split octonion form of field tensor
$F_{\mu\nu}'=\partial'_{\mu}B'_{\nu}-\partial'_{\nu}B'_{\mu}$ components
of which describe the electric and magnetic field in $T^{4}$- space.
Similarly, we obtain 

\begin{eqnarray}
D\overline{\emptyset} & = & F_{t}'\label{eq:70}\end{eqnarray}
where

\begin{eqnarray}
F_{t}' & = & \left(\begin{array}{cc}
0 & \overrightarrow{F_{t}}*\\
\overrightarrow{-F_{t}}* & 0\end{array}\right)=i\overrightarrow{\psi_{t}}.\label{eq:71}\end{eqnarray}
In the same manner, we get 

\begin{eqnarray}
\frac{1}{2}(\overline{D}\emptyset+D\overline{\emptyset}) & = & \left(\begin{array}{cc}
0 & \overrightarrow{H_{t}}\\
-\overrightarrow{H_{t}} & 0\end{array}\right)=\overrightarrow{H_{t}}\label{eq:72}\end{eqnarray}

\begin{eqnarray}
\frac{1}{2}i\,(\overline{D}\emptyset-D\overline{\emptyset}) & = & \left(\begin{array}{cc}
0 & \overrightarrow{E_{t}}\\
-\overrightarrow{E_{t}} & 0\end{array}\right)=\overrightarrow{E_{t}}\label{eq:73}\end{eqnarray}
where $\overrightarrow{E_{t}}$ and $\overrightarrow{H_{t}}$ are
the electric and magnetic fields of tachyons in $T^{4}-$space as
given by equations (\ref{eq:65},\ref{eq:66}).

\subsubsection*{(iii) DYONIC CASE}

In order to reformulate dyonic field equations in terms of split octonions
and its Zorn's vector matrix realization, we replace $\emptyset$
in equation (\ref{eq:42}) by the complex (generalized) four - potential
$\left\{ V_{\mu}\right\} $ of dyons \cite{key-36} in the following
form

\begin{eqnarray}
V & = & V_{4}u_{0}*+V_{4}u_{0}+V_{i}u_{i}*+V_{i}u_{i}=\left(\begin{array}{cc}
V_{4} & -\overrightarrow{V}\\
\overrightarrow{V} & V_{4}\end{array}\right)\label{eq:74}\end{eqnarray}
where $V_{4}$ and $\overrightarrow{V}$ are the temporal and spatial
components of generalized four - potential of dyons. These are complex
quantities and their real and imaginary components are electric and
magnetic constituents \cite{key-36}. Split octonion four - differential
operator is now defined as,

\begin{eqnarray}
D & =\partial_{4}u_{0}*+\partial_{4}u_{0}+\partial_{i}u_{i}*+\partial_{i}u_{i}= & \left(\begin{array}{cc}
\partial_{4} & -\overrightarrow{\nabla}\\
\overrightarrow{\nabla} & \partial_{4}\end{array}\right)\label{eq:75}\end{eqnarray}
Operating split octonion conjugate of four - differential operator
given by equation (\ref{eq:75}) on the equation (\ref{eq:74}) and
using to communication relations of split octonion units, we get

\begin{eqnarray}
\overline{D}V & = & \left(\begin{array}{cc}
\partial_{4}V_{4}+\overrightarrow{\nabla}.\overrightarrow{V} & -(\partial_{4}\overrightarrow{V}-\overrightarrow{\nabla}V_{4}-\overrightarrow{\nabla}\times\overrightarrow{V})\\
\partial_{4}\overrightarrow{V}-\overrightarrow{\nabla}V_{4}-\overrightarrow{\nabla}\times\overrightarrow{V} & \partial_{4}V_{4}+\overrightarrow{\nabla}.\overrightarrow{V}\end{array}\right)\label{eq:76}\end{eqnarray}
where

\begin{eqnarray}
\partial_{4}V_{4}+\overrightarrow{\nabla}.\overrightarrow{V} & = & (\partial_{\mu}V_{\mu})(u_{0}*+u_{0})=0.1=0\label{eq:77}\end{eqnarray}
and

\begin{eqnarray}
-\partial_{4}\overrightarrow{V}+\overrightarrow{\nabla}V_{4}+\overrightarrow{\nabla}\times\overrightarrow{V} & = & -i\psi*.\label{eq:78}\end{eqnarray}
Thus $V$ is the generalized split octonion potential of dyons and
$\overrightarrow{\psi}$ is the complex vector field with $\overrightarrow{E}$
and $\overrightarrow{H}$ as the generalized electromagnetic fields
of dyons. Consequently, equation (\ref{eq:76}) leads to the following
split octonion wave equation for dyonic field as

\begin{eqnarray}
\overline{D}V & = & G\label{eq:79}\end{eqnarray}
where

\begin{eqnarray}
G & = & \left(\begin{array}{cc}
0 & -\overrightarrow{G}\\
\overrightarrow{G} & 0\end{array}\right)=i\overrightarrow{\psi}*.\label{eq:80}\end{eqnarray}
It is to be noted from equations (\ref{eq:46}) to (\ref{eq:49})
that $f_{4},\,\overrightarrow{f\,},\overrightarrow{\, f}\,'$ and
$f_{4}'$ are the components of electric and magnetic fields described
in terms of two four-potentials in internal and external spaces. If
the two spaces are completely disjoint spaces then we get only $\overrightarrow{f}$,
in external four - space since $f_{4}=0$ due to Lorentz gauge condition
while $\overrightarrow{f}\,'$ and $f_{4\,}'$ do not occur. Equations
(\ref{eq:48}) and (\ref{eq:49}) shows that $\overrightarrow{f}$
and $\overrightarrow{f\,}'$ are made up from electric and magnetic
fields, which is the mixing of external and internal spaces. As such
equations (\ref{eq:48}) and (\ref{eq:49}) do not describe either
the usual electric and magnetic fields or the generalized fields of
dyons. Rather, they have the mixed behaviour, namely tachyonic dyons.

\subsection{{\normalsize FIELD EQUATIONS}}

Let us write the octonion wave equation in split representation as

\begin{eqnarray}
DF & = & J.\label{eq:81}\end{eqnarray}
Here $D$ and $F$ are defined by equations (\ref{eq:40}) and (\ref{eq:45})
in split octonionic form and $J$ is associated with eight dimensional
current source density in split octonion representation as follows,

\begin{eqnarray}
J & = & \left(\begin{array}{cc}
J_{4} & -\overrightarrow{J}\\
\overrightarrow{J} & J_{4}\end{array}\right).\label{eq:82}\end{eqnarray}
Now using equations (\ref{eq:40}) and (\ref{eq:45}) , we get

\begin{eqnarray}
DF & = & \left(\begin{array}{cc}
\partial_{4}f_{4}-\overrightarrow{\nabla}.\overrightarrow{f}\,' & -(\partial_{4}\overrightarrow{f}+\overrightarrow{\nabla}f_{4}'+\overrightarrow{\nabla}'\times f\,')\\
\partial'_{4}\overrightarrow{f}\,'+\overrightarrow{\nabla}'f_{4}+\overrightarrow{\nabla}\times\overrightarrow{f} & \partial'_{4}f_{4}'-\overrightarrow{\nabla}'.\overrightarrow{f}\end{array}\right).\label{eq:83}\end{eqnarray}
Comparing equations (\ref{eq:81}), (\ref{eq:82}) and (\ref{eq:83}),
we get

\begin{eqnarray}
J_{4} & = & \partial_{4}f_{4}-\overrightarrow{\nabla}.\overrightarrow{f}\,';\label{eq:84}\end{eqnarray}

\begin{eqnarray}
J_{4}' & = & \partial'_{4}f_{4}'-\overrightarrow{\nabla}'.\overrightarrow{f};\label{eq:85}\end{eqnarray}

\begin{eqnarray}
\overrightarrow{J} & = & \partial_{4}\overrightarrow{f}+\overrightarrow{\nabla}f_{4}'+\overrightarrow{\nabla}'\times f\,';\label{eq:86}\end{eqnarray}

\begin{eqnarray}
\overrightarrow{J}' & = & \partial'_{4}\overrightarrow{f}\,'+\overrightarrow{\nabla}'f_{4}+\overrightarrow{\nabla}\times\overrightarrow{f}.\label{eq:87}\end{eqnarray}
Let us discuss the following different cases of generalized field
equation.

\subsubsection*{(i) BRADYONIC CASE}

For the description of bradyons, let us substitute $B_{4}=A_{4},\,\overrightarrow{B}=\overrightarrow{A},\,\partial_{4}'=\partial_{4},\,\overrightarrow{\nabla}'=\overrightarrow{\nabla}\equiv\partial$
and $J_{4}'=J_{4},\,\overrightarrow{J}'=\overrightarrow{J}.$ Using
equations (\ref{eq:51}, \ref{eq:52}, \ref{eq:56}) we find that
equation (\ref{eq:83}) reduces to \begin{eqnarray}
DF & = & \left(\begin{array}{cc}
\overrightarrow{\nabla}.\mathcal{\overrightarrow{H}}+i\overrightarrow{\nabla}.\overrightarrow{E} & i\frac{\partial\mathcal{\overrightarrow{H}}}{\partial t}+i\overrightarrow{\nabla}\times\overrightarrow{E}-\frac{\partial\overrightarrow{E}}{\partial t}+\overrightarrow{\nabla}\times\mathcal{\overrightarrow{H}}\\
-i\frac{\partial\mathcal{\overrightarrow{H}}}{\partial t}-i\overrightarrow{\nabla}\times\overrightarrow{E}+\frac{\partial\overrightarrow{E}}{\partial t}-\overrightarrow{\nabla}\times\mathcal{\overrightarrow{H}} & \overrightarrow{\nabla}.\mathcal{\overrightarrow{H}}+i\overrightarrow{\nabla}.\overrightarrow{E}\end{array}\right)\label{eq:88}\end{eqnarray}
Comparing it with equations (\ref{eq:81} , \ref{eq:82}), we find
that equation (\ref{eq:88}) is analogous to the split octonionic
form of Maxwell's equation $F_{\mu\nu,\nu}=J_{\mu}$ which is the
covariant form of a set of following four differential equations i.e.

\begin{eqnarray*}
\overrightarrow{\nabla}.\overrightarrow{E}=\rho_{e}=-i\, J_{4}, & \overrightarrow{\nabla}.\overrightarrow{H}= & 0,\end{eqnarray*}

\begin{eqnarray}
\overrightarrow{\nabla}\times\overrightarrow{H}=-\overrightarrow{J}+\frac{\partial\overrightarrow{E}}{\partial t}, & \overrightarrow{\nabla}\times\overrightarrow{E}= & -\frac{\partial\overrightarrow{H}}{\partial t}.\label{eq:89}\end{eqnarray}
Hence the octonion field equation (\ref{eq:81}) is identical to the
Maxwell's field equation in compact and simple split octonion form.
Equation (\ref{eq:81}) may also be written in terms of potential
as \begin{eqnarray}
DF & =D(\overline{D}\emptyset) & =D\overline{D}\emptyset=\square\emptyset=J\label{eq:90}\end{eqnarray}
which is equivalent to split octonion form of covariant field equation
$\square V_{\mu}=J_{\mu}$ of classical electrodynamics. As such,
we have reformulated the classical electrodynamics in terms of compact,
simple and consistent representation of split octonion formulations
for the case of particles traveling slower than light namely bradyons.

\subsubsection*{(ii) TACHYONIC CASE}

In this case if we may substitute $A_{4}=B_{4},\,\overrightarrow{A}=\overrightarrow{B},\,\partial_{4}=\partial_{4}',\,\overrightarrow{\nabla}'=\overrightarrow{\nabla}\equiv\partial$
and $J_{4}=J_{4}',\,\overrightarrow{J}=\overrightarrow{J}\,'.$ Hence,
using equations (\ref{eq:61}) and (\ref{eq:64}), we get

\begin{eqnarray}
DF & '= & \left(\begin{array}{cc}
J_{4}' & -\overrightarrow{J}\,'\\
\overrightarrow{J} & J_{4}\end{array}\right)=J\,'.\label{eq:91}\end{eqnarray}
This equation is the split octonionic form of Maxwell's equation $F''_{\mu\nu,\nu}=J_{\mu}'$
in $T^{4}\,$- space where we have used the following pairs of Maxwell's
equation for tachyons {[}\cite{key-25,key-26}]

\begin{eqnarray*}
\overrightarrow{\nabla}'.\overrightarrow{E}_{t}=-\rho_{0}, & \overrightarrow{\nabla}'.\overrightarrow{H_{t}}= & 0,\end{eqnarray*}

\begin{eqnarray}
\overrightarrow{\nabla}'\times\overrightarrow{H_{t}}=-\overrightarrow{J}\,'+\frac{\partial\overrightarrow{E_{t}}}{\partial t}, & \overrightarrow{\nabla}'\times\overrightarrow{E_{t}}= & -\frac{\partial\overrightarrow{H_{t}}}{\partial t}.\label{eq:92}\end{eqnarray}
Hence equation (\ref{eq:90}) is visualized as the field equation
of tachyons in $T^{4}\,$- space in compact, simple and consistent
split octonionic formulation. It has already been concluded \cite{key-23,key-24,key-25}
that $T^{4}\,$- space for tachyons plays the same role as bradyons
do in $R^{4}\,$- space.

\subsubsection*{(iii) DYONIC CASE}

Here, we consider $\left\{ A_{\mu}\right\} $ and $\left\{ B_{\mu}\right\} $
as electric and magnetic four - potentials described in external space
(i.e. $R^{4}\,$-space) only, $D$ is defined by equation (\ref{eq:75})
and $\emptyset$ is replaced by $V$ the generalized four- potential
of dyons given by equation (\ref{eq:74}) . Subsequently, on using
equations (\ref{eq:75}), (\ref{eq:76}) and (\ref{eq:79}), we get

\begin{eqnarray}
D(\overline{D}V) & = & DG=J\label{eq:93}\end{eqnarray}
where

\begin{eqnarray}
J & =J_{4}u_{0}*+J_{4}u_{0}+J_{i}u_{i}*+J_{i}u_{i}= & \left(\begin{array}{cc}
J_{4} & -\overrightarrow{J}\\
\overrightarrow{J} & J_{4}\end{array}\right)\label{eq:94}\end{eqnarray}
is the split octonion form of generalized four - current associated
with dyons. Accordingly, we get

\begin{eqnarray}
D(\overline{D}V) & =(D\overline{D})V=DG & =J\label{eq:95}\end{eqnarray}
or equivalently

\begin{eqnarray}
\square V & = & J\label{eq:96}\end{eqnarray}
where

\begin{eqnarray}
\square & = & (\overline{D}D)=(D\overline{D})=\partial_{4}^{2}+\nabla^{2}=\partial_{i}\partial_{i}.\label{eq:97}\end{eqnarray}
Equations (\ref{eq:93}) and (\ref{eq:95}) are the split octonion
equivalents of Generalized Dirac-Maxwell's (GDM) equation of dyons.
Hence we may describe split octonion wave equation as the generalized
field equation of dyons in compact, simple and consistent manner.
Moreover, equation (\ref{eq:93}) and (\ref{eq:95}) reproduce the
field equations corresponding to the dynamics of electric charge (magnetic
monopole) in the absence of magnetic (electric) charge on the dyons.
Equation (\ref{eq:95}) is the thus split octonion equivalent of covariant
field equations $\square V_{\mu}=J_{\mu}$ of generalized fields of
dyons described earlier \cite{key-26}.

\subsection{CURRENT EQUATION}

Let us discuss the octonion wave equation as the octonion current
differential equations in eight - dimension as

\begin{eqnarray}
DJ & = & S.\label{eq:98}\end{eqnarray}
Here $D$ and $J$ are defined earlier equations (\ref{eq:41}) and
(\ref{eq:82}) in their split form and $S=\square F$ is defined as

\begin{eqnarray}
S & = & \left(\begin{array}{cc}
S_{4} & -\overrightarrow{S}\\
\overrightarrow{S}' & S_{4}'\end{array}\right).\label{eq:99}\end{eqnarray}
Now using equations (\ref{eq:41}) and (\ref{eq:46}) and equation
(\ref{eq:82}) , we get

\begin{eqnarray}
DJ & = & \left(\begin{array}{cc}
\partial_{4}J_{4}-\overrightarrow{\nabla}.\overrightarrow{J}\,' & -(\partial_{4}\overrightarrow{J}+\overrightarrow{\nabla}J_{4}'+\overrightarrow{\nabla}'\times\overrightarrow{J}\,')\\
\partial'_{4}\overrightarrow{J}\,'+\overrightarrow{\nabla}'J_{4}+\overrightarrow{\nabla}\times\overrightarrow{J} & \partial'_{4}J_{4}'-\overrightarrow{\nabla}'.\overrightarrow{J}\end{array}\right).\label{eq:100}\end{eqnarray}
Now using equations (\ref{eq:98}), (\ref{eq:99}) and (\ref{eq:100})
, we get the following set of equations

\begin{eqnarray}
S_{4} & = & \partial_{4}J_{4}-\overrightarrow{\nabla}.\overrightarrow{J}\,';\label{eq:101}\end{eqnarray}

\begin{eqnarray}
S_{4}' & = & \partial'_{4}J_{4}'-\overrightarrow{\nabla}'.\overrightarrow{J};\label{eq:102}\end{eqnarray}

\begin{eqnarray}
\overrightarrow{S} & = & \partial_{4}\overrightarrow{J}+\overrightarrow{\nabla}J_{4}'+\overrightarrow{\nabla}'\times\overrightarrow{J}\,';\label{eq:103}\end{eqnarray}

\begin{eqnarray}
\overrightarrow{S}' & = & \partial'_{4}\overrightarrow{J}\,'+\overrightarrow{\nabla}'J_{4}+\overrightarrow{\nabla}\times\overrightarrow{J};\label{eq:104}\end{eqnarray}
where $S$ is the new octonion variable parameter obtained from the
operation of differential operator to current.Hence we may obtain
a new kind of field equation in terms of new parameter $S$ and accordingly
we may analyze the different cases for bradyons, tachyons and dyons.

\textbf{Acknowledgment}- The work is supported by Uttarakhand Council
of Science and Technology, Dehradun. One of us OPSN is thankful to
Chinese Academy of Sciences and Third world Academy of Sciences for
awarding him CAS-TWAS visiting scholar fellowship to pursue a research
program in China. He is also grateful to Professor Tianjun Li for
his hospitality at Institute of Theoretical Physics, Beijing, China.

\end{document}